\documentclass[english,prb,floatfix,twocolumn,superscriptaddress]{revtex4-2}
\usepackage{ae,aecompl}
\usepackage[T1]{fontenc}
\usepackage[utf8]{inputenc}
\usepackage{babel}
\usepackage{float}
\usepackage{amsmath}
\usepackage{amssymb}
\usepackage{graphicx}
\usepackage{wasysym}
\usepackage{color}
\usepackage{physics}
\usepackage{dsfont}
\usepackage[normalem]{ulem}
\usepackage[unicode=true,bookmarks=true,bookmarksnumbered=false,bookmarksopen=false,breaklinks=false,pdfborder={0 0 1},backref=false,colorlinks=true,linkcolor=magenta,citecolor=blue]{hyperref}

\usepackage{changes}

\makeatletter
\newcommand\resetchangescolor[1]{%
  \setkeys{Changes@definechangesauthor}{color=#1}%
  \expandafter%
  \let\csname Changes@AuthorColor\endcsname=\Changes@definechangesauthor@color%
  \colorlet{Changes@Color}{\@nameuse{Changes@AuthorColor}}%
}

\begin{document}
\global\long\def\ket#1{\left|#1\right\rangle }%
\global\long\def\bra#1{\left\langle #1\right|}%
\global\long\def\braket#1#2{\langle#1|#2\rangle}%
\global\long\def\expectation#1#2#3{\langle#1|#2|#3\rangle}%
\global\long\def\average#1{\langle#1\rangle}%

\title{Effects of reservoir squeezing on the amplification of quantum correlation}
\author{Zhaorui Peng}
\affiliation{Department of Physics, Zhejiang Normal University, Jinhua, Zhejiang 321004, China}
\author{Lucas C. Céleri\href{https://orcid.org/0000-0001-5120-8176}{\includegraphics[scale=0.05]{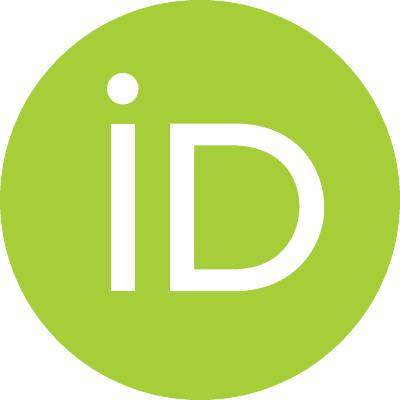}}}
\email{lucas@qpequi.com}
\affiliation{QPequi Group, Institute of Physics, Federal University of Goi\'{a}s, 74.690-900, Goi\^{a}nia, Brazil}
\author{Abdul Basit}
\email{abasitphy@gmail.com}
\affiliation{Department of Physics, Zhejiang Normal University, Jinhua, Zhejiang 321004, China}
\affiliation{Department of Applied Physics and Astronomy, College of Sciences, University of Sharjah, Sharjah 27272, United Arab Emirates}
\author{Gao Xianlong\href{https://orcid.org/0000-0001-6914-3163}{\includegraphics[scale=0.05]{orcidid.pdf}}}
\email{gaoxl@zjnu.edu.cn}
\affiliation{Department of Physics, Zhejiang Normal University, Jinhua, Zhejiang 321004, China}
\date{\today}

\begin{abstract}
The dynamics of quantum discord is studied in a system of two identical noninteracting qubits coupled to a common squeezed vacuum bath through non-demolition interactions. We concern on how reservoir squeezing influences the dynamical behaviors of quantum discord when both qubits are initially prepared in $X$-type states. We find that the critical time $t_c$ exhibits the sudden change of quantum discord, which is of great significance for the quantum discord amplification. Furthermore, depending on the initial parameters of the system, we numerically calculate the interval when the critical time $t_c$ is finite or infinite. For the finite critical time $t_c$, we show that the squeezing phase of the bath can prolong the critical time $t_c$ while the squeezing strength exhibits the opposite effect. For infinite critical time $t_c$, even if there is no sudden transition point, reservoir squeezing still has an effect on the amplification of quantum discord, and the time to reach steady-state quantum discord can be changed by adjusting the squeezing parameters. Finally, we investigate the quantum speed limit time for a two-qubit system under squeezed reservoir, and find that the quantum speed limit time can be reduced via the adjustment of the squeezing parameters and the initial parameters. Remarkably, in the short time limit, reservoir squeezing has an obvious influence on the degree of amplification of quantum discord. Our study presents a promising approach to controlling the amplification of quantum correlation.
\end{abstract}

\maketitle 

\section{\label{Sec:1}Introduction}

The core of quantum technologies is the generation of deterministic quantum resources. It is well known that quantum entanglement acts as the most important resources in quantum information processing~\cite{Yu2004,Yu2009,Bellomo2007,de Moraes Neto2022,Mintert2005,Chitambar2019,Horodecki2009,Wootters1998,Vidal2002}. However, the entanglement may disappear completely after a finite time, a phenomenon called sudden death of entanglement~\cite{Yu2004,Yu2009}. Thus, it is necessary to take into account quantum correlations beyond entanglement~\cite{Modi2010,Oppenheim2002,Zurek2003,Vedral2003,Bera2018,Datta2008,Henderson2001,Groisman2005,Maziero2009}. Compared to other measures, quantum discord is considered as the most suitable resource to quantify classical and nonclassical correlations~\cite{Kaszlikowski2008,Ferraro2010}, which was first introduced by Ollivier and Zurek~\cite{Ollivier2001}. Quantum discord has the non-negative property and is able to detect quantum correlations not only in entangled states but also in separable states. Moreover, the dynamics of quantum discord has been investigated both in theory and in experiment~\cite{Lanyon2013,Fanchini2010,Maziero2010}. Notably, the analytical expressions for classical correlation and quantum discord are difficult to obtain except for two-qubit Bell diagonal states and a seven-parameter family of two-qubit $X \text {-type}$ states~\cite{Luo2008,Ali2010,Chen2011}. Motivated by the study on bipartite systems, quantum discord has also been investigated for the multipartite system~\cite{Li2021,Rulli2011}. Overall, there exist two fundamental forms of quantum discord, measurement-based discord and distance-based discord. Some studies have investigated the phenomenon of frozen and time-invariant quantum discord for Markovian and non-Markovian environments~\cite{Addis2015,Haikka2013,ABasit2021}. In addition, quantum discord has been used in the field of quantum phase transition~\cite{Maziero2012}, quantum metrology and quantum electrodynamics.

One of the main difficulties in manipulating quantum devices is that dissipation or decoherence tends to occur due to the interaction between the system and its environment, in which dissipation occurs when energy is transferred from the system to its surrounding while there is no energy exchange in decoherence~\cite{Streltsov2017}. Unfortunately, dissipation or decoherence results in the loss of important quantum features. Hence, understanding the impact of the environment on quantum correlation is crucial and essential. In several previous studies, quantum discord exhibits more robustness than the entanglement when measuring the quantum correlation~\cite{Werlang2009}. The behavior of the quantum discord under decoherence is a highly active area of research~\cite{Xu2011,Zhang2014,Karpat2011}. In the case of the Markovian environment, the open system dynamics exhibits an abrupt change from the classical to the quantum decoherence regime~\cite{Mazzola2010}. On the other hand, for a system under local nondissipative non-Markovian channels, the memory effects of the environment result in multiple transitions~\cite{mazzola2010}.

In the previously mentioned investigations, it was common to regard the reservoir (environment) initially as being in either a thermal or a vacuum state. However, as reservoir engineering technology has progressed~\cite{Poyatos1996}, it is now possible to create a non-thermal state for open quantum systems. The squeezed bath can be created by operating a Josephson parametric amplifier in a superconducting circuit or manipulating a trapped impurity in a double-well potential~\cite{Murch2013,Nir2011}. Furthermore, the use of non-thermal bath techniques yielded some remarkable results in the field of theoretical research. In the paper~\cite{Akhtar2023}, the author has investigated the Wigner function of the photon-added squeezed-vacuum states~\cite{Akhtar2022,Dey2020,Biswas2007}, which garnered attention in quantum metrology. For the non-Markovian dynamics of the open system, squeezed bath plays an effective regulatory role~\cite{Ablimit2023,MAli2010,He2019}. In addition, the squeezed thermal bath has also been studied in quantum thermodynamics~\cite{You2018,Agarwalla2017,Klaers2017,Ro2014,Assis2021,Assis2020}. Besides, squeezed thermal and vacuum baths make a significant impact on various phenomena such as entanglement sudden-death~\cite{Hernandez2008,MAli2010}, violation of Leggett-Garg-type inequalities~\cite{Naikoo2019}, enhancing the lifetime of the cat state~\cite{Drummond2020}, and several others~\cite{Banerjee2007,Drummond2004,Wang2019}. The above-mentioned research has sparked our interest in the effects of squeezed thermal/vacuum baths on quantum correlation.

In quantum information processing, the evolution of the system of interest can change under different baths. The minimum evolution time between two distinguishable quantum states is called quantum speed limit (QSL) time which is established from the energy-time uncertainty principle~\cite{Mandelstam1945,Margolus1998}. Determination of QSL time is significant in the areas of quantum computation~\cite{Ashhab2012}, quantum state transmission~\cite{Nie2021}, quantum thermodynamics~\cite{Mukhopadhyay2018,D. P. Pires2021}, quantum metrology~\cite{Giovannetti2011}, etc. More recently, the QSL time for the dephasing model in a squeezed bath has been studied~\cite{He2019,Du2021}. In the literature, the QSL time can be reduced by adjusting the squeezing parameters of the bath. Inspired by these results, we will investigate the QSL time of a two-qubit system in a common squeezed reservoir. Recent works in the connection of quantum correlation and QSL time reveal that QSL time could be used to analyze the dynamics of quantum correlations~\cite{Paulson2022,Tiwari2023,Paulson2205}. In this work, we address the question how fast steady-state quantum correlation can be reached in a system of two identical non-interacting qubits under different squeezed reservoirs.

In our previous research~\cite{basit2021}, we have exposed the impacts of the squeezing of the two spatially separated baths on the appearance of frozen discord and the sudden transition between classical and quantum decoherence. Furthermore, in the other paper~\cite{Basit2023,Basit2021}, we examined the dynamics of one-norm geometric quantum correlations and their classical counterparts in a two-qubit system. Remarkably, we have found that the emergence of a pointer-state basis \text { (quantum-to-classical transition) } can also be delayed by adequately adjusting the squeezing parameters of one common bath. However, the role of reservoir squeezing on the creation of quantum correlation is still not clear; in this paper, we focus on the amplification of quantum discord between two identical qubits, which are subjected to a common squeezed vacuum bath. It is worth noting that the phenomenon of quantum discord amplification exists by selecting appropriate initial parameters. 

Referring to the study on the amplification of quantum discord~\cite{Yuan2010}, it has been confirmed that when a two-qubit system is subjected to a common Ohmic environment, the amplification or protection of quantum discord can occur during the time evolution. Specifically, researchers have discovered that in the case of two identical qubits, stable amplification of the quantum discord exists. In addition, it is found that the quantum phase transition in the cavity-BEC system is the physical mechanism of sensitive quantum discord amplification~\cite{Yuan2013}. 

The organization of this article is as follows. In Sect.~\ref{Sec:2}, we introduce the physical dephasing model and its solution. In Sect.~\ref{Sec:3}, we study the quantum discord dynamical behaviors of the two identical qubits in a common squeezed vacuum bath for $X \text {-type}$ initial states. In Sect.~\ref{Sec:4}, we study numerically the effects of reservoir squeezing on the quantum discord amplification when the critical time $t_c$ is infinite and finite, respectively. In Sect.~\ref{Sec:5}, we study the QSL time for the two-qubit system in a common squeezed vacuum bath. Finally, we conclude in Sect.~\ref{Sec:6}.

\section{\label{Sec:2} Decoherence model}

We start by considering the pure dephasing model~\cite{Gardiner2004,Agarwal2012,Breuer2002}, which is composed by two identical qubits coupled to a common squeezed environment. The total Hamiltonian can be written as $H =H_\textrm{S}+H_\textrm{B}+H_\textrm{SB}$, where
\begin{equation*}
H_\textrm{S} =\frac{1}{2} \sigma_A^z \omega_0+\frac{1}{2} \sigma_B^z \omega_0
\end{equation*}
and 
\begin{equation*}
H_\textrm{B} = \sum_{k} b_{k}^{\dagger} b_{k} \omega_k
\end{equation*}
are the system and the environment free Hamiltonians, respectively, while the interaction between the qubits and the environment is given by
\begin{equation*}
H_\textrm{SB} =  \sum_{\alpha=\textrm{A},\textrm{B}}\sum_{k}\sigma_\alpha^z\left(g_{k} b_{k}^{\dagger}+g_{k}^{*} b_{k}\right).
\end{equation*}
In these equations, $\sigma_{\alpha}^{z}$ represents the $z$ component of the Pauli operators for the qubit $\alpha$. We represent by $|e\rangle_{\alpha}$ ($|g\rangle_{\alpha}$) the excited (ground) state of the qubit $\alpha$, whose transition frequency is $\omega_0$. Furthermore, $b_k^{\dagger}$ and $b_k$ are the creation and annihilation operators for $k$-th mode of the environment, associated with frequency $\omega_k$. Finally, $g_{k}$ is the coupling strength between the qubits and the $k$-th mode of the environment. For convenience, we choose the couplings to be real.

The dynamics of the system can be obtained by considering first the case of a single qubit. The time evolution can be straightforwardly generalized to the case of the two-qubit system. The Hamiltonian for the single-qubit system is
\begin{equation}
\label{2}
H=\frac{1}{2} \sigma_z \omega_0+\sum_{k} b_{k}^{\dagger} b_{k} \omega_k+\sum_{k} \sigma_z\left(g_{k} b_{k}^{\dagger}+g_{k}^* b_{k}\right).
\end{equation}

Now, we assume that the system and the environment are initially uncorrelated, $\rho_\textrm{SB}(0)=\rho_\textrm{S}(0) \otimes \rho_\textrm{B}(0)$, and consider that the bath is initially in the squeezed thermal state given by
\begin{equation}
\label{3}
\rho_\textrm{B}(0)=\prod_k \hat{S}_{k}\left(r, \theta\right) \rho^{k}_{\textrm{th}} \hat{S}_{k}^{\dagger}\left(r, \theta\right),
\end{equation}
with
\begin{equation}
\label{4}
\hat{S}_{k}\left(r, \theta\right)=\exp \left(\frac{1}{2} \zeta^* b_k^2-\frac{1}{2} \zeta b_k^{\dagger^2}\right),
\end{equation}
being the squeezing operator for the mode $k$ with $\zeta=r e^{i \theta}$, where $r$ and $\theta$ indicate the squeezing strength and squeezing phase, respectively, while 
\begin{equation}
\label{5}
\rho^{k}_{\textrm{th}}=\frac{\exp \left(-\beta \omega_k b_k^{\dagger} b_k\right)}{\operatorname{Tr} \exp \left(-\beta \omega_k b_k^{\dagger} b_k\right)},
\end{equation}
with $\beta$ being the inverse temperature ($k_B=1$), represents the thermal state. It is convenient to consider the time evolution of the system in the interaction picture. The system dynamics is governed by
\begin{equation*}
U(t) = \exp \left[\sum_{k}\sigma_z\left(\xi_k(t) b_{k}^{\dagger}-\xi_k^*(t) b_{k}\right)\right],
\end{equation*}
with $\xi_k(t)=g_k(1-e^{i \omega_k t}) / \omega_k$. Since $\left[\sigma_z, H\right]=0$, implying that the populations of the reduced density matrix,  $\rho_\textrm{S}(t)=\Tr_\textrm{B}\left[U(t) \rho_\textrm{SB}(0) U^{\dagger}(t)\right]$ remain constant in time. The coupling with the bath destroys quantum coherence, reflected in the exponential decay of the off-diagonal terms of the reduced density matrix in the basis $\{|g\rangle,|e\rangle\}$~\cite{Breuer2002}
\begin{equation}
\label{7}
\rho_\textrm{S}(t)=\left(\begin{array}{cc}
\rho_\textrm{S}^{ee} & \rho_\textrm{S}^{eg} e^{-\Gamma(t)} \\
\rho_\textrm{S}^{ge} e^{-\Gamma(t)} & \rho_\textrm{S}^{gg} 
\end{array}\right),
\end{equation}
with
\begin{equation}\label{8}
e^{-\Gamma(t)}=\operatorname{tr}_{\textrm{B}}\left(\rho_{\textrm{B}}(0) \exp \left\{2 \sum_k\left[\xi_k(t) b_{k}^{\dagger}-\xi_k^*(t) b_{k}\right]\right\}\right).
\end{equation}
Therefore, the dephasing factor $\Gamma(t)$ in Eq.~(\ref{7}) can be written as~\cite{He2019}
\begin{equation}
\label{9}
\Gamma(t)=\sum_k \frac{1}{2}\left|\eta_k(t)\right|^2 \operatorname{coth} \frac{\beta \omega_k}{2},
\end{equation}
with $\eta_k(t)=2 \xi_k(t) \cosh r+2 \xi_k^*(t) e^{i \theta} \sinh r$. Then substituting $\xi_k(t)=g_k(1-e^{i \omega_k t}) / \omega_k$ into the above expression, the dephasing factor takes the form
\begin{equation}
\label{10}
\begin{aligned}
\Gamma(t) & =\sum_k \frac{4\left|g_k\right|^2}{\omega_k^2}\left(1-\cos \omega_k t\right) \operatorname{coth} \frac{\beta \omega_k}{2} \\
& \times\left\{\cosh 2 r-\sinh 2 r \cos \left(\omega_k t-\theta\right)\right\},
\end{aligned}
\end{equation}
Then in the continuum limit, $\sum_k \rightarrow \int_0^{\infty} \mathrm{d} \omega J(\omega)$, where $J(\omega)=2 \pi \sum_k\left|g_k\right|^2 \delta\left(\omega-\omega_k\right)$ is the spectral density of the environment, we obtain
\begin{equation}
\label{11}
\begin{aligned}
\Gamma(t) & =4 \int_0^{\infty} \frac{d \omega}{2\pi} J(\omega) \operatorname{coth} \frac{\beta \omega}{2}\frac{1-\cos \omega t}{\omega^2}\\
& \times\{\cosh 2 r -\sinh 2 r \cos (\omega t-\theta)\}.
\end{aligned}
\end{equation}
In our work, we assume the Ohmic environment with the spectral density $J(\omega)=\omega e^{-\omega / \omega_c}/2$~\cite{Chakravarty1984}. $\omega_c$ is the cutoff frequency. Furthermore, the analytical solution to Eq.~(\ref{11}) can be obtained in the zero temperature limit as
\begin{equation}
\label{12}
\begin{aligned}
\Gamma(t) & =\frac{1}{2\pi}\left\{A(t) \cosh 2 r-\sinh 2 r\left(B(t) \cos \theta\right.\right. \\
& \left.\left.+C(t) \sin \theta\right)\right\},
\end{aligned}
\end{equation}
with the time-dependant coefficients $A(t)=\ln \left[1+\tau^2\right]$, $B(t)=\ln \left[1+4 \tau^2\right]^{ 1/2} - \ln[1+\tau^2]$, and $C(t)=2 \arctan \tau-\arctan 2 \tau$, with $\tau=\omega_c t$. Equation~\eqref{12} shows that the action of the squeezed thermal environment is significantly different from the environment without squeezing, even at zero temperature~\cite{Breuer2002}. Therefore, from here on we focus on the evolution of the two-qubit system in the squeezed reservoir at zero temperature.

\section{\label{Sec:3}Classical and quantum correlations}

In this section, we focus on studying the time evolution of the quantum discord for the two identical qubits under the action of a common environment at the squeezed vacuum state. The total amount of classical and quantum correlations in the two-qubit quantum system are measured by the quantum mutual information
\begin{equation}
\label{13}
\mathcal{I}\left(\rho_{\textrm{AB}}\right)=S\left(\rho_\textrm{A}\right)+S\left(\rho_\textrm{B}\right)-S\left(\rho_\textrm{AB}\right),
\end{equation}
where $S(\rho)=-\Tr\left\{\rho \log _2 \rho\right\}$ is the von Neumann entropy. For a bipartite state $\rho_\textrm{AB}$, the quantum discord $\mathcal{Q}\left(\rho_\textrm{AB}\right)$ is defined as~\cite{Ollivier2001,Henderson2001},
\begin{equation}
\label{14}
\mathcal{Q}\left(\rho_\textrm{AB}\right) \equiv \mathcal{I}\left(\rho_\textrm{AB}\right)-\mathcal{C}\left(\rho_\textrm{AB}\right),
\end{equation}
where the classical correlation $\mathcal{C}\left(\rho_{A B}\right)$ is given by,
\begin{equation}
\label{15}
\begin{aligned}
\mathcal{C}\left(\rho_\textrm{AB}\right) & =\max _{\left\{\Pi_k^\textrm{B}\right\}}\left\{S\left(\rho_\textrm{A}\right)-\sum_k p_k S\left(\rho_{\textrm{A} \mid k}\right)\right\}\\
& =S\left(\rho_\textrm{A}\right)-\min _{ \left\{\Pi_k^\textrm{B}\right\}}\left[\sum_k p_k S\left(\rho_{\textrm{A} \mid k}\right)\right].
\end{aligned}
\end{equation}
Here, $\left\{\Pi_k^\textrm{B}\right\}$ is a set of projective measurements performed on subsystem $\textrm{B}$, and $p_k \rho_{\textrm{A} \mid k}= \Tr_\textrm{B}\left[\left(\mathds{1}_\textrm{A} \otimes \Pi_k^\textrm{B}\right) \rho_\textrm{AB}\left(\mathds{1}_\textrm{A} \otimes \Pi_k^\textrm{B}\right)\right]$ is the reduced density matrix of the subsystem $\textrm{A}$ after measurements on $\textrm{B}$ with probability $p_k=\Tr_\textrm{AB}\left[\left(\mathds{1}_\textrm{A} \otimes \Pi_k^\textrm{B}\right) \rho_\textrm{AB}\left(\mathds{1}_\textrm{A} \otimes \Pi_k^\textrm{B}\right)\right]$, where $\mathds{1}_\textrm{A}$ is the identity operator for the subsystem $\textrm{A}$. Generally, to finding the analytical expression of quantum discord is very difficult. For simplicity, we take the initial state of the composite system as a class of states with maximally mixed marginals described by the $X \text {-type}$ density matrix~\cite{Luo2008,Ali2010},
\begin{equation}
\label{16}
\begin{aligned}
\rho_\textrm{AB}(0) & =\frac{1}{4}\left(\mathds{1}_{\textrm{AB}}+\sum_{i=1}^3 c_i \sigma_\textrm{A}^i \otimes \sigma_\textrm{B}^i\right) \\
& =\frac{1}{4}\left(\begin{array}{cccc}
1+c_3 & 0 & 0 & c_1-c_2 \\
0 & 1-c_3 & c_1+c_2 & 0 \\
0 & c_1+c_2 & 1-c_3 & 0 \\
c_1-c_2 & 0 & 0 & 1+c_3
\end{array}\right),
\end{aligned}
\end{equation}
where $\mathds{1}_\textrm{AB}$ is the identity operator of the the two qubits and the real numbers $c_i\left(0 \leqslant\left|c_i\right| \leqslant 1\right)$ satisfy the conditions that the density matrix $\rho_\textrm{AB}(0)$ is a positive matrix and the trace of it is unit. This class of states includes the Werner states $\left(\left|c_1\right|=\left|c_2\right|=\left|c_3\right|=c\right)$ and Bell states $\left(\left|c_1\right|=\left|c_2\right|=\left|c_3\right|=1\right)$.

We are interested in the time evolution of a pure dephasing model,  where a two-qubit system is coupled with a common squeezed vacuum bath.  In the Hilbert space spanned by the two-qubit product state basis $\{|e e\rangle,|e g\rangle,|g e\rangle,|g g\rangle\}$, we can obtain the density matrix $\rho_\textrm{AB}(t)$ of two identical qubits initially prepared in state~\eqref{16} as~\cite{Palma1996}
\begin{equation}
\label{17}
\rho_\textrm{AB}(t)=\frac{1}{4}\left(\begin{array}{cccc}
1+c_3 & 0 & 0 & \alpha(t)\\
0 & 1-c_3 & c_1+c_2 & 0 \\
0 & c_1+c_2 & 1-c_3 & 0 \\
\alpha(t)& 0 & 0 & 1+c_3
\end{array}\right),
\end{equation}
where $\alpha(t)=\left(c_1-c_2\right) e^{-4 \Gamma(t)}$, and $\Gamma(t)$ is the dephasing factor defined in Eq.~(\ref{12}). The eigenvalues of the density matrix $\rho_\textrm{AB}(t)$ are
\begin{equation}
\label{18}
\begin{aligned}
\mu_{1,2} & =\frac{1}{4}\left[1+c_3 \mp \alpha(t)\right], \\
\mu_{3,4} & =\frac{1}{4}\left[1-c_3 \mp\left(c_1+c_2\right)\right].
\end{aligned}
\end{equation}

Then the analytical expression of mutual information can be written as 
\begin{equation}
\label{19}
\mathcal{I}\left[\rho_\textrm{AB}(t)\right]=2+\sum_{n=1}^4 \mu_n \log _2 \mu_n.
\end{equation}

We now need to compute the classical correlations given in Eq.~\eqref{15}. This can be performed by writing the complete set of orthogonal projectors as $\Pi_k=\left|\vartheta_k\right\rangle\left\langle\vartheta_k\right|$, with $k=1, 2$, where
\begin{equation}
\label{20}
\begin{aligned}
\left|\vartheta_1\right\rangle & =\cos \vartheta|g\rangle+e^{i \phi} \sin \vartheta|e\rangle, \\
\left|\vartheta_2\right\rangle & =e^{-i \phi} \sin \vartheta|g\rangle-\cos \vartheta|e\rangle,
\end{aligned}
\end{equation}
with $0 \leqslant \vartheta \leqslant \pi / 2 \text { and } 0 \leqslant \phi \leqslant 2 \pi$ representing the usual polar coordinates. By performing the maximization over these parameters, we obtain~\cite{Maziero2009}
\begin{equation}
\label{22}
\mathcal{C}\left(\rho_\textrm{AB}(t)\right)=\sum_{j=1}^2 \frac{1+(-1)^j \chi(t)}{2} \log _2\left[1+(-1)^j \chi(t)\right],
\end{equation}
where
\begin{equation}
\label{23}
\chi(t)=\max \left\{\left|c_3\right|,(|\alpha(t)|+|c_1+c_2|) / 2\right\}.
\end{equation}
Then the quantum discord between the two identical qubits can be written as,
\begin{equation}
\label{24}
\mathcal{Q}\left(\rho_\textrm{AB}(t)\right)=2+\sum_{n=1}^4 \mu_n \log _2 \mu_n-\mathcal{C}\left(\rho_\textrm{AB}(t)\right).
\end{equation}

\begin{figure}[h]
\includegraphics[width=3.2 in]{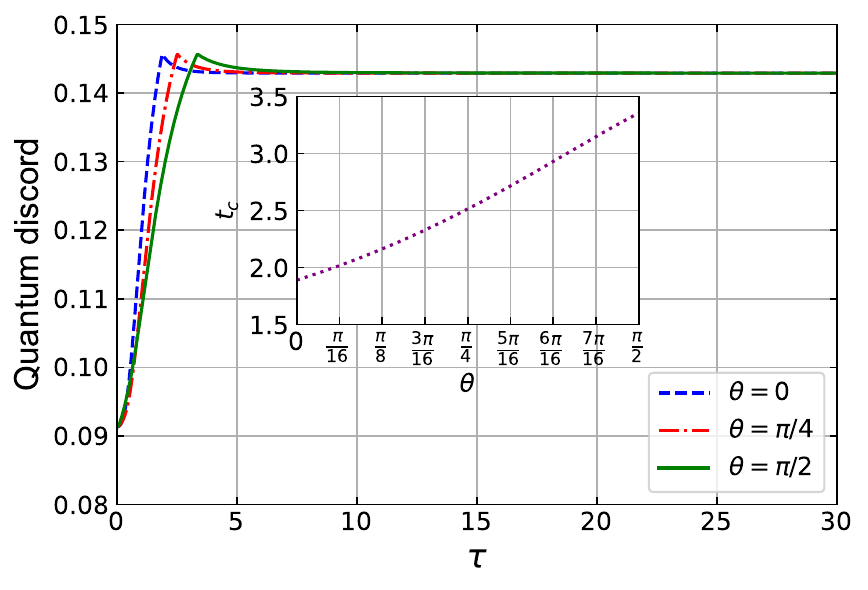}
\includegraphics[width=3.2 in]{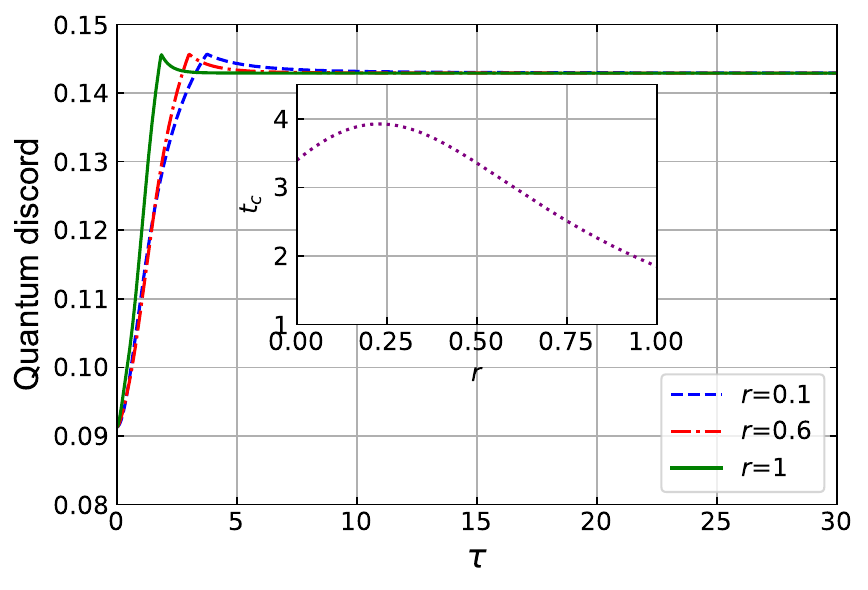}
\caption{Time evolution of the quantum discord of two qubits subjected to a common squeezed thermal bath at zero temperature for different values of the squeezing phase $\theta$ (upper panel) and  the squeezing strength $r$ (lower panel). Inset (a): The critical time as a function of $\theta$. Inset (b): The critical time as a function of $r$.} 
\label{Time evolution of the quantum discord finite}
\end{figure}
\begin{figure}[h]
\includegraphics[width=3.2 in]{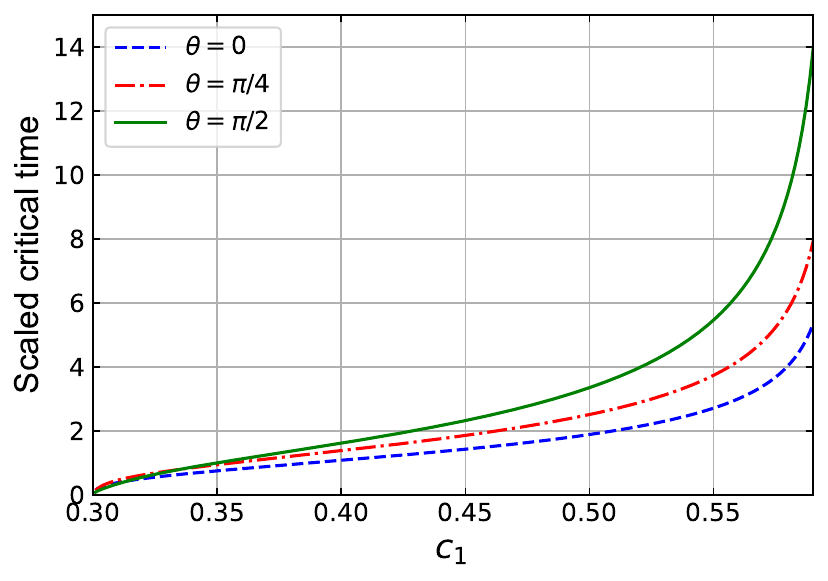}
\includegraphics[width=3.2 in]{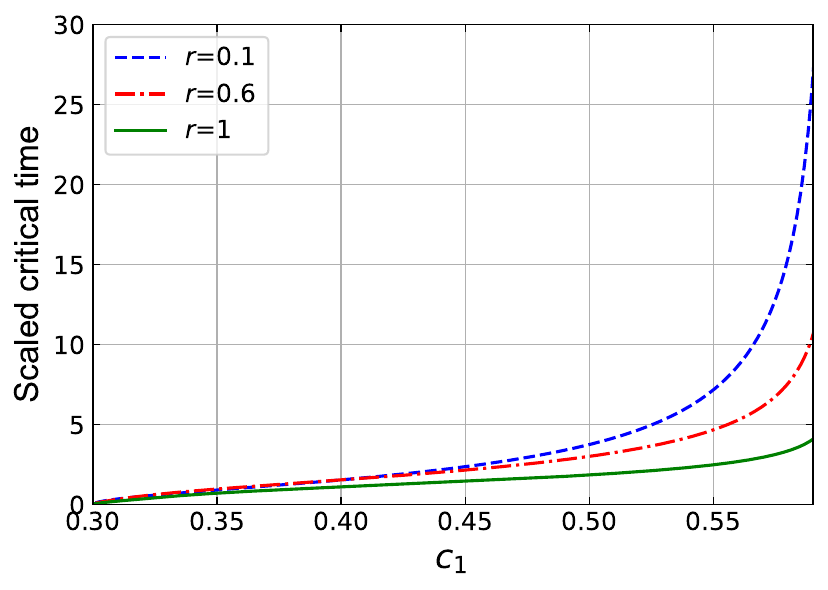}
\caption{The scaled critical time $\omega_c t_c$ as a function of the $c_1$ for the common squeezed thermal bath at zero temperature with different values of the squeezing phase $\theta$ (upper panel) and  the squeezing strength $r$ (lower panel).} 
\label{The scaled critical time}
\end{figure}
The dynamical behaviour of the quantum discord is shown in Fig.~\ref{Time evolution of the quantum discord finite}. It is clear that quantum discord increases in time, and by changing the squeezing parameters we change the speed of amplification. This indicates that squeezing results in the amplification of quantum correlations. We can also see that quantum discord increases until it reaches a maximum, after which the dynamics changes. $\mathcal{Q}$ approaches a constant value which is independent of the squeezing parameters. However, there is an important phenomenon that happens when we change the reservoir, the time $t_{c}$ at which the maximum is achieved changes. Although changes in phase result in an increase in $t_{c}$, the increase in squeezing strength decreases $t_{c}$. In order to better understand this effect, we now study the sudden change phenomenon of quantum correlations~\cite{Maziero2009}.

\section{\label{Sec:4}Sudden change of quantum discord}

As we saw in the last section, the dynamics of the quantum correlations changes at certain instants of time $t_{c}$. These are the critical times identified in Ref.~\cite{Maziero2009}. The purpose of this section is to study how $t_{c}$ depends on the squeezing parameters.

To start, we note that $\alpha(t)$ is a decaying function of time. Therefore, the function $\chi(t)$ defined in Eq.~(\ref{23}) reveals that there may exist a critical time $t_c$ at which quantum discord exhibits a sudden change in the dynamics~\cite{Maziero2009}. From Eq.~\eqref{23} we can see that this critical time obeys the following condition
\begin{equation}
\label{25}
\frac{\left|\alpha\left(t_c\right)\right|+\left|c_1+c_2 \right|}{2}=\left|c_3\right|.
\end{equation}

It is evident that the critical time $t_c$ depends on the initial state and on the characteristics of the reservoir. In the last section we saw that quantum discord exhibits different dynamics before and after the critical time. Let us then analytically calculate the quantum discord for different time intervals.

Before the critical time, i.e., $t \in\left[0, t_c\right)$, we have $\chi(t)=(|\alpha(t)|+|c_1+c_2|) / 2$. The classical correlation reads
\begin{equation}
\label{26}
\mathcal{C}\left(\rho_\textrm{AB}(t)\right)=\sum_{j=1}^2 \frac{1+(-1)^j \Omega(t)}{2} \log _2\left[1+(-1)^j \Omega(t)\right],
\end{equation}
where $\Omega(t)=(|\alpha(t)|+|c_1+c_2|) / 2$. The quantum discord can be obtained from Eqs.~\eqref{18} and~(\ref{24}).

In the regime of $t>t_c$, we have $\chi(t)=\left|c_3\right|$. The classical correlation is a constant
\begin{equation}
\label{28}
\mathcal{C}\left(\rho_\textrm{AB}(t)\right)=\sum_{j=1}^2 \frac{1+(-1)^j \left|c_3\right|}{2} \log _2\left[1+(-1)^j \left|c_3\right|\right].
\end{equation}
However, the quantum discord still depends on time (see Eq.~\eqref{18}). But now, since the classical correlations do not change, $\mathcal{Q}$ approaches stability very fast, as can be seem from Fig.~\ref{Time evolution of the quantum discord finite}. 

Now, depending on the behaviour of $\alpha(t)$, $t_c$ can be finite or infinite. In order to understand this, we will treat each case separately. 
\begin{figure*}[tbp]
\includegraphics[width=\textwidth]{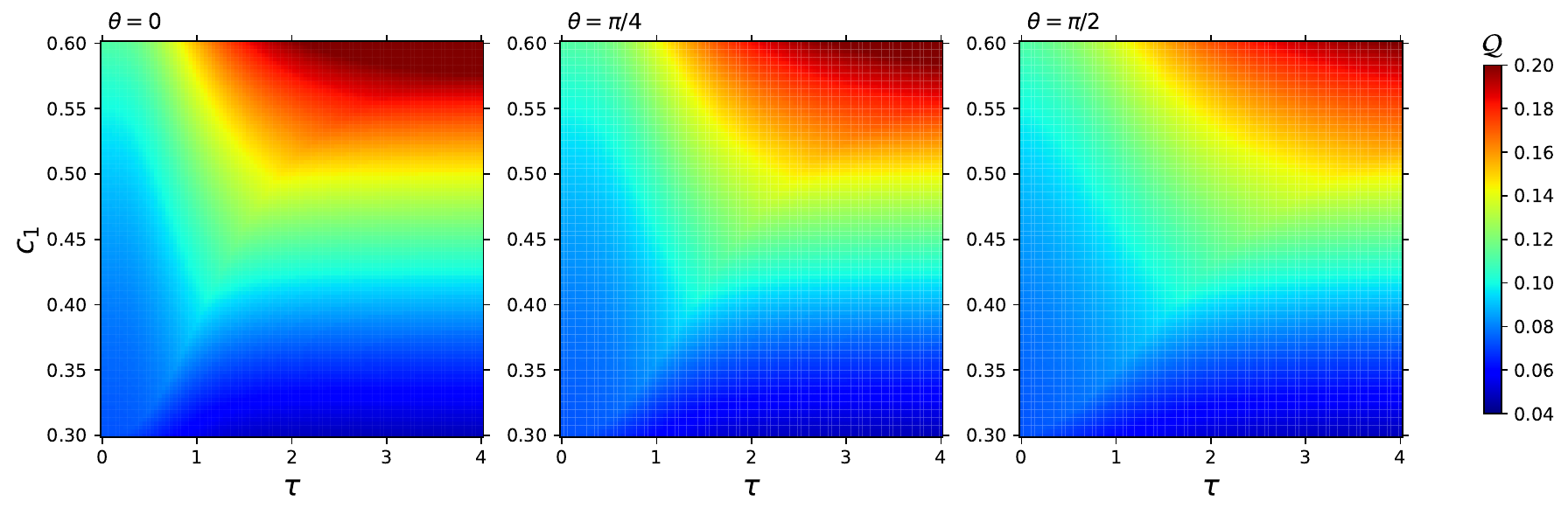}
\includegraphics[width=\textwidth]{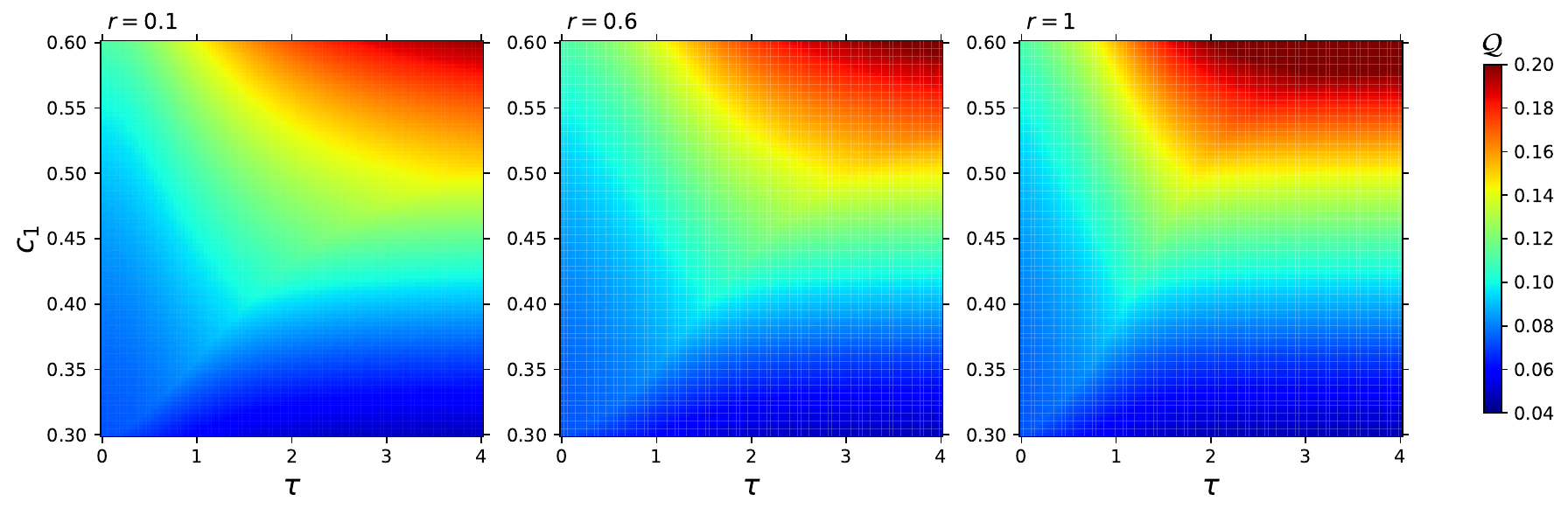}
\caption{The phase diagram of the quantum discord $\mathcal{Q}$ as functions of the time $\tau$ and the initial-state $c_1$ for different values of the squeezing parameters. The other parameters are set as $c_2=0, c_3=0.3$.}
\label{The quantum discord finite}
\end{figure*}
\subsection{The case of finite critical time}

First, it should be noted that the analytical expressions for the critical time $t_c$ are difficult to obtain for the squeezed vacuum. However, solutions to Eq.~\eqref{25} can always be numerically found. In Ref.~\cite{Yuan2010}, the analytical result for $t_c$ is obtained for some initial states with $0<c_1 / 2 \leqslant c_3<c_1 \leqslant 2 / 3 \text { and } c_2=0$ in a vacuum bath without squeezing. For different values of the squeezing phase $\theta = \lbrace 0,\pi / 4,\pi / 2\rbrace$ we find that the critical time $t_c$ is finite for the initial states with $0.3<c_1<0.6$, $c_2=0$, $c_3=0.3$, and squeezing strength $r=0.5$. In Fig.~\ref{The scaled critical time} we plot the critical time $t_c$ as a function of $c_1$. As already anticipated in Fig.~\ref{Time evolution of the quantum discord finite} the critical time scales differently regarding the phase and the strength of the squeezing. Moreover, we can see from Fig.~\ref{The scaled critical time} that $t_{c}$ increases with $c_{1}$, which means that it takes more time for the dynamics to reach stabilization.

Figure~\ref{The scaled critical time} demonstrate that increasing the phase $\theta$ leads to the increase of $t_{c}$ for all values of $c_{1}$, while the opposite occurs with respect to the squeezing strength. Therefore, by choosing a small value of $\theta$ and a large value of $r$ we can reach stabilization very fast.  Since the maximum value of quantum discord is not affected by squeezing, reaching stabilization faster can be good for applications in quantum information processing, since quantum correlations can be employed as a resource. 

It is worth noting that according to Eq.~(\ref{11}), we know that the dephasing factor $\Gamma(t)$ is a monotonically increasing function, which causes $\alpha(t)\rightarrow 0$ as time approaches infinite. In other words, the squeezing effect will be lost in the long-time evolution. In order to discuss in more detail the influence of the squeezing on the quantum discord at the short time limit, we present the phase diagram for different squeezing parameters in Fig.~\ref{The quantum discord finite}. The value of the quantum discord is shown to depend strongly on $c_1$ and the squeezing parameters, as expected. Fig.~\ref{The quantum discord finite} clearly indicates that the quantum discord is not amplified when the initial parameter $c_1\lesssim 0.42$. However, the amplification of quantum discord
becomes more apparent with increasing $c_{1}$.

\begin{figure}[tbp]
\includegraphics[width=3.2 in]{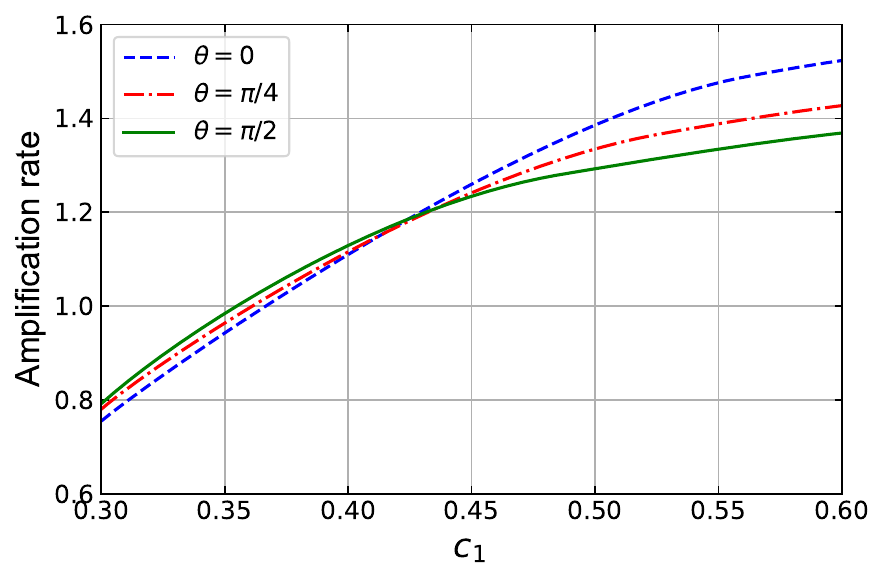}
\includegraphics[width=3.2 in]{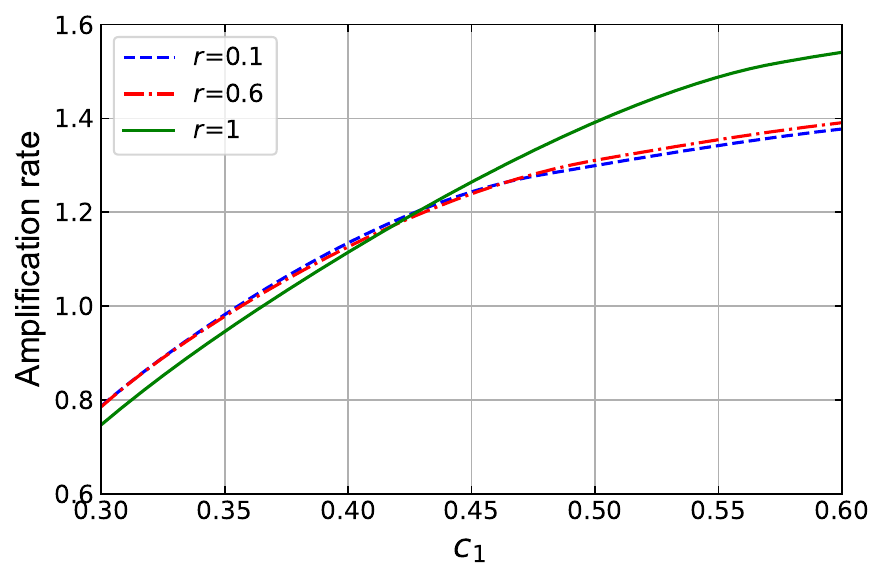}
\caption{The amplification rate of average quantum discord $R=\overline{\mathcal{Q}\left(\rho\left(3\right)\right)}/\mathcal{Q}\left(\rho\left(0\right)\right)$ of two qubits is plotted as a function of the initial-state parameter $c_1$ for the common squeezed thermal bath at zero temperature with (a) the different squeezing phase $\theta$ and (b) the squeezing strength $r$. Other parameters are set as $c_2=0, c_3=0.3$.} 
\label{The amplification rate of quantum discord finite}
\end{figure}

In order to better explore the properties of amplification of short-time quantum discord in a squeezed vacuum, we define the amplification rate of the average quantum discord in the time interval $\tau \in[0,3]$ as $R=\overline{\mathcal{Q}\left(\rho\left(3\right)\right)}/\mathcal{Q}\left(\rho\left(0\right)\right)$, where $\overline{\mathcal{Q}\left(\rho\left(\tau^{\prime}\right)\right)}=\int_0^{\tau^{\prime}} \mathcal{Q}(\rho(\tau)) d \tau$. In Fig.~\ref{The amplification rate of quantum discord finite} we plot the amplification rate $R$ as a function of the initial-state parameter $c_1$. Fig.~\ref{The amplification rate of quantum discord finite} indicates that the larger $c_1$, the less apparent the increase of the $R$. And different squeezing parameters will lead to varying degrees of increase in the amplification rate. Furthermore, there exists an intersection point for the different squeezing phase/strength in Fig.~\ref{The amplification rate of quantum discord finite}(a) and Fig.~\ref{The amplification rate of quantum discord finite}(b), respectively. More specifically, the coordinates of the intersection points  are (0.421, 1.176) and (0.436, 1.219), respectively. Notably, the squeezing effect is not significant before the intersection point. However, the rate $R$ increases uniformly with increasing squeezing phase, while there is a significant increase as the squeezing strength $r$ approaches 1 after the intersection point. In other words, the decrease in the slope of the amplification rate function is influenced by the squeezing parameters. In addition, Fig.~\ref{The amplification rate of quantum discord finite} shows that the initial prepared quantum discord can not be amplified in the whole initial parameter range.

Let us move now to the case of infinite $t_{c}$.

\subsection{The case of infinite critical time}

\begin{figure}[tbp]
\includegraphics[width=3.2 in]{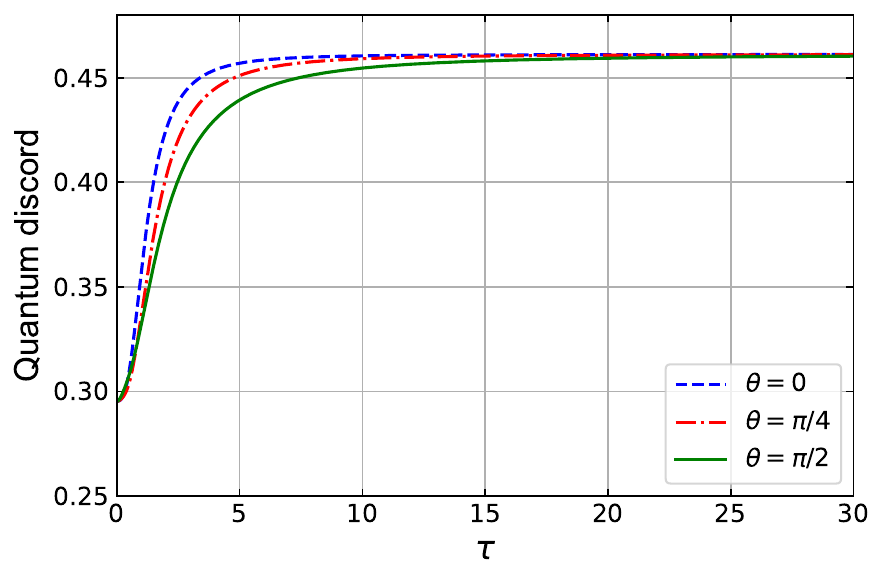}
\includegraphics[width=3.2 in]{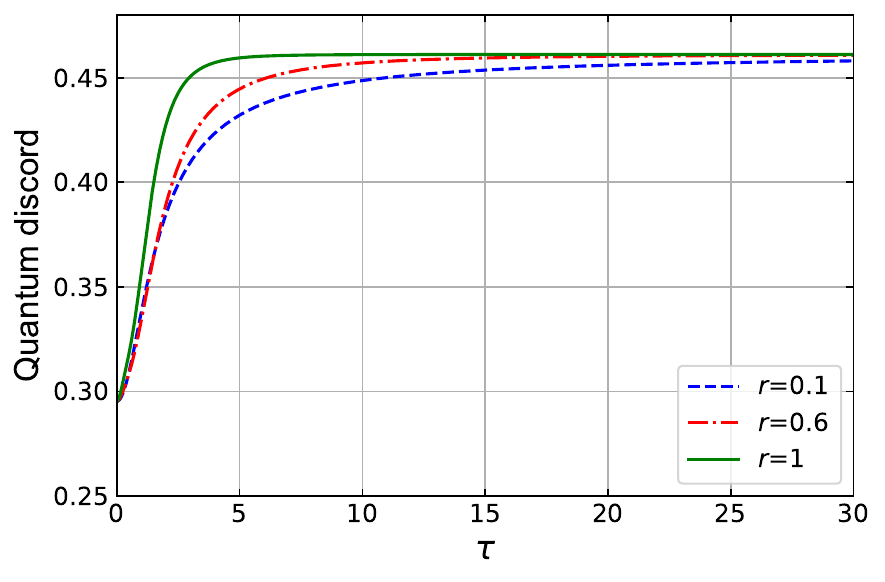}
\caption{Time evolution of the quantum discord of two qubits subjected to a common squeezed thermal bath at zero temperature for different values of the squeezing phase $\theta$ (upper panel) and  the squeezing strength $r$ (lower panel). Other parameters are set as $c_1=0.9, c_2=0.6, c_3=-0.6$.} 
\label{Time evolution of the quantum discord infinite}
\end{figure}

After extensive exploration, we find that the critical time $t_c$ is infinite for the initial states with $0.6<c_1<1$, $c_2=0.6$, $c_3=-0.6$. In Fig.~\ref{Time evolution of the quantum discord infinite}, we plot the dynamical evolution of the quantum discord at zero temperature for different values of the squeezing phase parameters. Again, the value of the steady-state quantum discord is not affected squeezing. The same behavior on the dependence of the time to achieve the steady-state on the squeezing is observed here, although without a sudden change. If we increase the squeezing phase of the bath, the time to reach steady-state quantum discord will be delayed. However, it takes a shorter time to achieve steady-state quantum discord by increasing the squeezing strength. As shown in Ref.~\cite{Zhang2014}, the time to reach steady-state quantum discord can also be changed via bang-bang pulses with a finite period. Thus, the squeezed vacuum bath technique can be used as a new scheme to regulate the amplification of quantum discord in addition to dynamical decoupling.

\begin{figure*}[h]
\includegraphics[width=\textwidth]{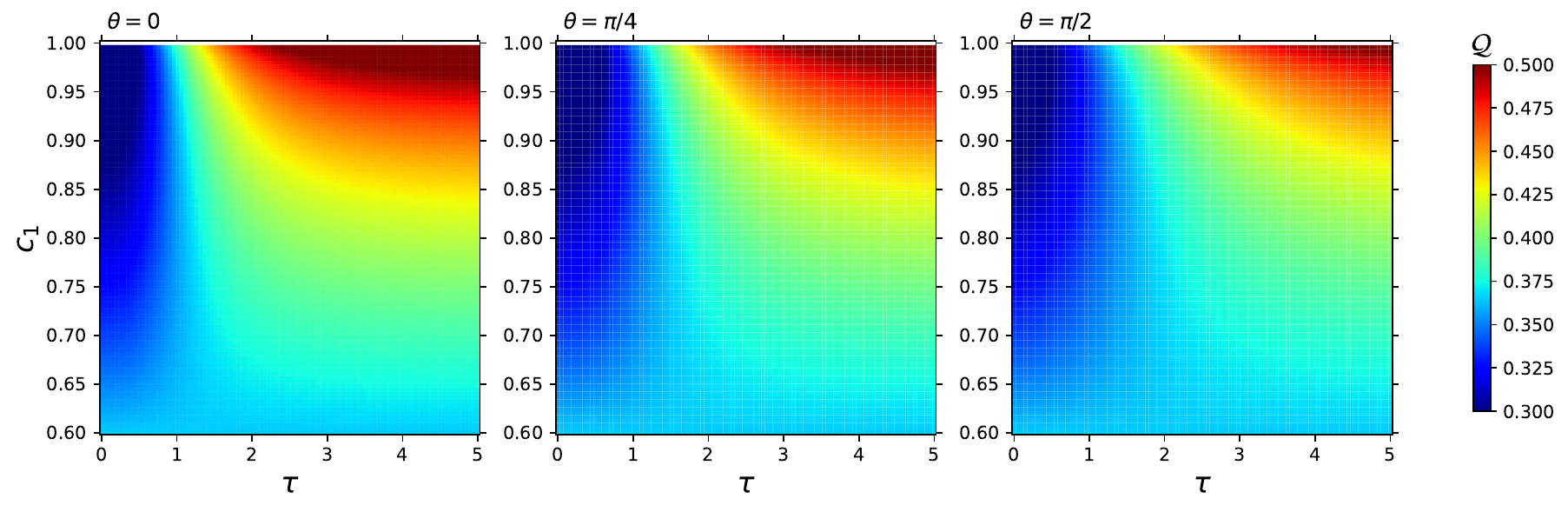}
\includegraphics[width=\textwidth]{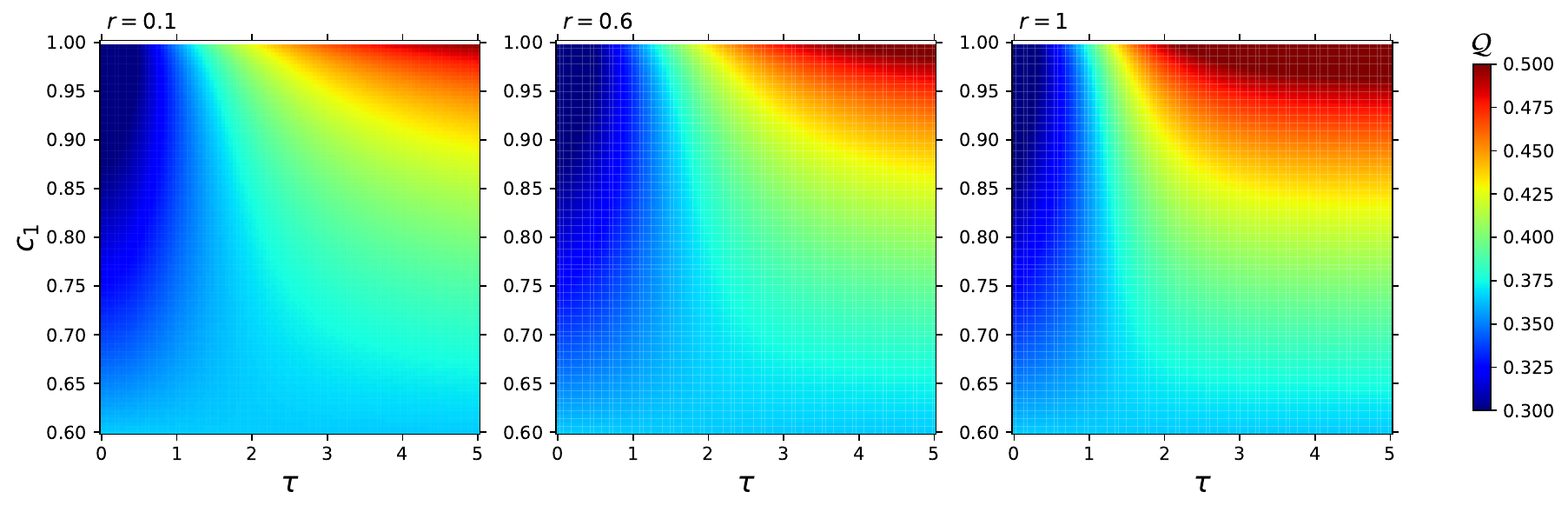}
\caption{The phase diagram of the quantum discord $\mathcal{Q}$ as functions of the time $\tau$ and the initial-state parameter $c_1$ for different values of the squeezing parameters. Other parameters are set as $c_2=0.6, c_3=-0.6$.}
\label{The quantum discord infinite}
\end{figure*}

For completeness, in Fig.~\ref{The quantum discord infinite}, we plot the phase diagram of quantum discord in the case of infinite critical time. Obviously, quantum discord can be amplified with increasing time $\tau$ and $c_1$. Remarkably, we can obtain a similar result that the rate of reaching steady-state quantum discord can be changed by adjusting the squeezing parameters in Fig.~\ref{The quantum discord finite}.

\begin{figure}[tbp]
\includegraphics[width=3.2 in]{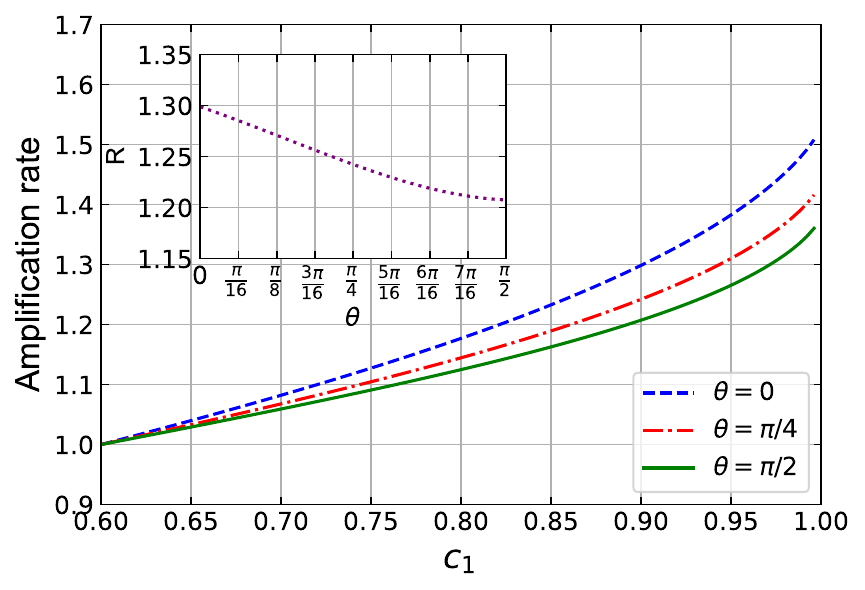}
\includegraphics[width=3.2 in]{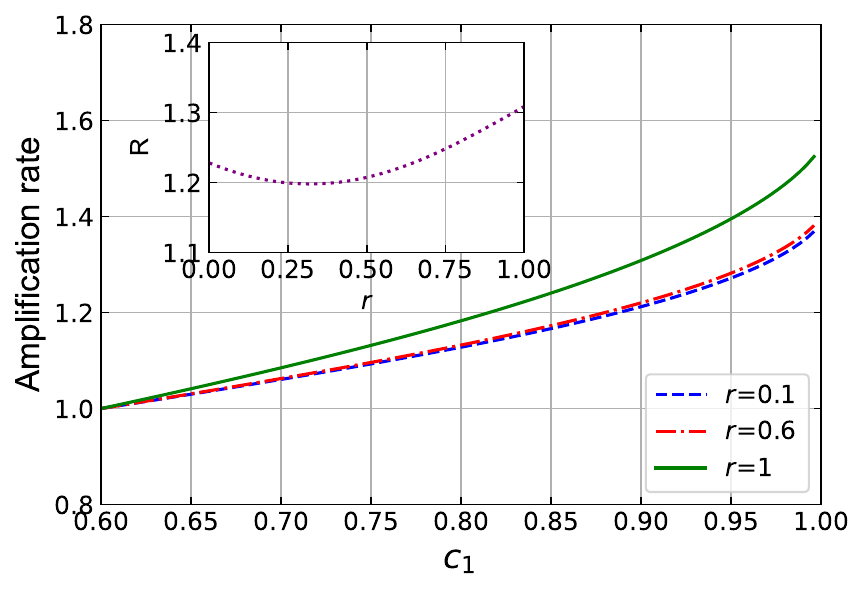}
\caption{The amplification rate of average quantum discord $R=\overline{\mathcal{Q}\left(\rho\left(3\right)\right)}/\mathcal{Q}\left(\rho\left(0\right)\right)$ of two qubits is plotted as a function of the initial-state parameter $c_1$ for the common squeezed thermal bath at zero temperature with the different value of (a) the squeezing phase and (b) the squeezing strength. Inset (a): The amplification rate of the quantum discord $R$ is plotted as a function of squeezing phase with $c_1=0.9, c_2=0.6, c_3=-0.6$. Inset (b): The amplification rate of the quantum discord $R$ is plotted as a function of squeezing strength with $c_1=0.9, c_2=0.6, c_3=-0.6$. Other parameters are set as $c_2=0.6, c_3=-0.6$.}
\label{The amplification rate of quantum discord infinite}
\end{figure}

In Fig.~\ref{The amplification rate of quantum discord infinite} we show the amplification rate $R$ as a function of the initial state parameter $c_1$. It shows that, for all initial state parameters $c_1$ within the set value range, the quantum discord of two identical qubits in a squeezed vacuum bath can be amplified, and the amplification rate $R$ increases with increasing initial parameter $c_1$. The same amount of adjustment of the squeezing phase will cause the amplification rate $R$ to change uniformly, while the squeezing strength will cause a large change of the amplification rate $R$ only at large values. Generally, Fig.~\ref{The amplification rate of quantum discord infinite} indicates that the squeezing strength has a positive effect on short-time quantum discord amplification while the squeezing phase has a negative effect. In particular, we plot the amplification rate with respect to the squeezing phase/strength when $c_1=0.9, c_2=0.6, c_3=-0.6$. Interestingly, we find that the amplification rate decreases as the squeezing phase increases. However, the amplification rate does not monotonically increase with the increase of squeezing strength. According to Ref.~\cite{Yuan2010}, we know that the quantum coherence effect has triggered the quantum discord amplification for the two identical qubits in a common bath. Remarkably, the squeezed vacuum bath will not change the basic trend of the quantum discord amplification, but can adjust the degree of the amplification.

Since we have identified the role of the squeezing parameters on the dynamical behaviour of quantum correlations, we now provide an analysis of the quantum speed limit, which will tell us how fast the amplification process can be achieved.

\section{\label{Sec:5}Quantum speed limit}

\begin{figure}[h]
\includegraphics[width=3.2 in]{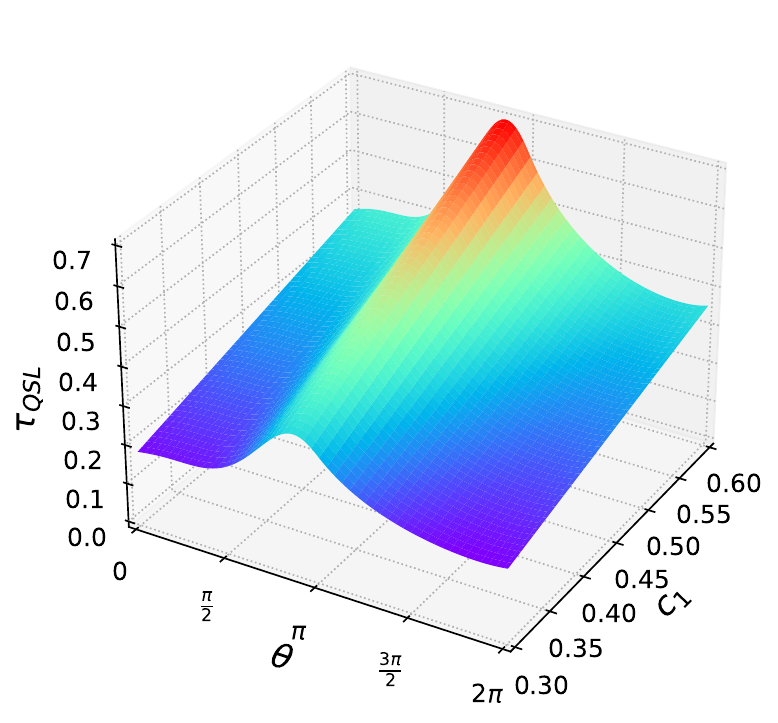}
\includegraphics[width=3.2 in]{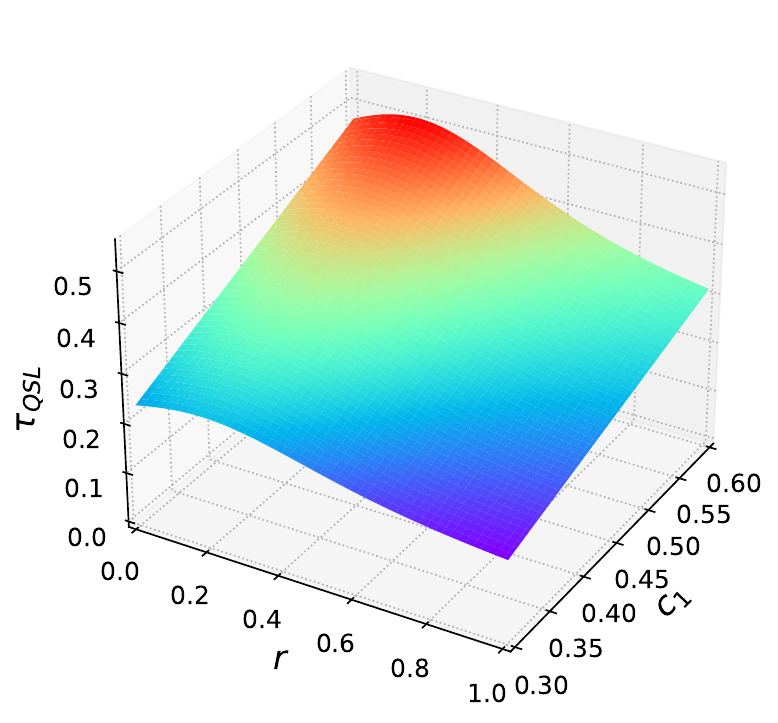}
\caption{The QSL time of two qubits is plotted
as functions of initial-state parameter $c_1$ and the squeezing parameter, which is (a) the squeezing strength $r=0.5$ and (b) the squeezing phase $\theta=\pi / 2$. The driving time is chosen as $\tau=1$. Other parameters are set as $c_2=0$, $c_3=0.3$.} 
\label{QSL}
\end{figure}

In this section, we consider the QSL time for the dephasing model under a squeezed vacuum bath. The geometric approach in quantum information theory is generally applied to estimate Mandelstam-Tamm (MT) and Margolus-Levitin (ML)-type bounds in QSL time~\cite{Deffner2013,Deffner2017}. Geodesic length $\mathcal{L}$ represents a bona fide measure of distinguishability in quantum state space. However, there are only two Riemannian metrics admitting known geodesics, which are quantum Fisher information and the Wigner-Yanase skew information. The more widely used of the two metrics is the quantum Fisher information metric~\cite{Pires2016},
\begin{equation}
\label{A1}
\mathcal{L}^{Q F}(\rho_0, \rho_\tau)=\arccos [\sqrt{F(\rho_0, \rho_\tau)}],
\end{equation}
where $F\left(\rho_0, \rho_\tau\right)=\left(\operatorname{tr}\left[\sqrt{\sqrt{\rho_0} \rho_\tau \sqrt{\rho_0}}\right]\right)^2$ is the Uhlmann fidelity. Firstly, we consider the QSL time in a closed quantum system. In Ref.~\cite{Pires2016}, the authors have provided the Mandelstam-Tamm type bound on the rate of quantum unitary evolution, 
\begin{equation}
\label{C2}
\tau \geq \frac{\hbar}{\overline{\Delta E_\tau}} \mathcal{L}^{Q F}\left(\rho_0, \rho_\tau\right),
\end{equation}
where we have introduced the time averaged variance of the Hamiltonian, $\overline{\Delta E_\tau}:=\tau^{-1} \int_0^\tau d t \sqrt{\left\langle H_t^2\right\rangle-\left\langle H_t\right\rangle^2}$. It is worth noting that Eq.~(\ref{C2}) applies to arbitrary initial and final mixed states and arbitrary time-dependent Hamiltonians. Now we consider the Wigner-Yanase information metric,
\begin{equation}
\label{C3}
\mathcal{L}^{W Y}\left(\rho_0, \rho_\tau\right)=\arccos [A\left(\rho_0, \rho_\tau\right)],
\end{equation}
where $A(\rho_0, \rho_\tau)=\operatorname{Tr}(\sqrt{\rho_0} \sqrt{\rho_\tau})$ is the quantum affinity. The QSL time emerging from the Wigner-Yanase information metric can be represented as
\begin{equation}
\label{C4}
\tau\geq \frac{1}{\sqrt{2}} \frac{\hbar}{\overline{\Delta E_\tau}} \mathcal{L}^{W Y}\left(\rho_0, \rho_\tau\right).
\end{equation}
When the initial and final states commute, the corresponding fidelity and affinity have the same value, which means that the QSL time corresponding to the quantum Fisher information metric is tighter than the one corresponding to the Wigner-Yanase information metric. However, the latter provides the tighter lower bound in the open system dynamics~\cite{Pires2016}. Therefore, we make use of the Hellinger angle $\mathcal{L}^{W Y}\left(\rho_0, \rho_\tau\right)$ between the initial and final states of the quantum system to obtain the QSL time under nonunitary dynamics. In general, the dynamics of an open quantum system are governed by the time-dependent master equation $\dot{\rho}_t=L\left(\rho_t\right)$ with the Liouvillian super-operator $L\left(\rho_t\right)$. Suppose that the initial state of the system is a pure state $\rho_0=\left|\psi_0\right\rangle\left\langle\psi_0\right|$, employing the von Neumann trace inequality and the Cauchy-Schwarz inequality for operators, the MT-type and ML-type bounds of the QSL time is given as follows.
\begin{equation}
\label{C5}
\tau \geq \tau_{\mathrm{QSL}}=\max \left\{\frac{1}{\Lambda_\tau^{\mathrm{op}}}, \frac{1}{\Lambda_\tau^{\mathrm{tr}}}, \frac{1}{\Lambda_\tau^{\mathrm{hs}}}\right\} \sin ^2 (\mathcal{L}^{W Y}\left(\rho_0, \rho_\tau\right)),
\end{equation}
where $\Lambda=1 / \tau \int_0^\tau d t\left\|L\left(\rho_t\right)\right\|$, and $\Lambda_\tau^{\mathrm{op}}, \Lambda_\tau^{\mathrm{tr}} \text {, and } \Lambda_\tau^{\mathrm{hs}}$ are the operator, Hilbert-Schmidt and trace norms, respectively. Due to the fact that the matrix norms satisfy the inequality $\|\cdot\|_{\mathrm{tr}} \geq\|\cdot\|_{\mathrm{hs}} \geq\|\cdot\|_{\mathrm{op}}$, it is clear that the QSL time $\tau_{\mathrm{QSL}}$ with respect to the operator norm is the tightest for open quantum systems. However, we take the initial state of the
two-qubit system as a class of states with maximally mixed marginals. In the case of  mixed initial states, since it should be treated by purification in a sufficiently enlarged Hilbert space, the statistical distance measured by the Bures angle is in general not suitable. Here we adopt the function of relative purity as the generalized Bloch angle to define the QSL time, which is given by~\cite{Campaioli2018}
\begin{equation}
\label{C6}
\Theta\left(\rho_0, \rho_t\right)=\arccos \left(\sqrt{\frac{\operatorname{tr}\left[\rho_0 \rho_t\right]}{\operatorname{tr}\left[\rho_0^2\right]}}\right).
\end{equation}
Based on the metric introduced in Eq.~(\ref{C6}), a lower bound of the non-unitary evolution time that is applicable to mixed initial states was proposed~\cite{ZhangY2014,Wu2018,Campo2013}:
\begin{equation}
\label{C7}
\tau_{\mathrm{QSL}}=\frac{1}{\Lambda_\tau^{\mathrm{op}}} \sin ^2\left[\Theta\left(\rho_0, \rho_\tau\right)\right] \operatorname{tr}\left[\rho_0^2\right].
\end{equation}
It should be noted that the operator norm $ 
\left\|L\left(\rho_t\right)\right\|_{\mathrm{op}}$ is identical to the largest singular value of $L\left(\rho_t\right)$. The non-unitary generator of the reduced dynamics of the system is $L\left(\rho_t\right)=\frac{\gamma(t)}{2}\left(\sigma_z \rho_t \sigma_z-\rho_t\right)$, with $\gamma(t)$ being the dephasing rate, i.e., the derivative of dephasing factor $\Gamma(t)$. Thus, the denominator in Eq.~(\ref{C7}) can be written as 
\begin{equation}
\label{C8}
\Lambda_\tau^{\mathrm{op}}=\frac{1}{\tau} \int_0^\tau \mathrm{d} t\left(c_1-c_2\right)\left|\gamma(t) e^{-4\Gamma(t)}\right|.
\end{equation}
In Fig.~\ref{QSL}, the QSL time has been shown as a function of the initial parameter $c_1$ and the squeezing parameters $r$ and $\theta$, respectively. Here, the initial correlation parameters are set as $c_2=0$, $c_3=0.3$ and the actual driving time is chosen as $\tau=1$. From Fig.~\ref{QSL}(a), we see that the quantum speed limit will be tighter when the initial parameter $c_1$ is larger. As shown in Ref.~\cite{S. Wu2014}, the dependence on the initial state is an important factor in signaling the acceleration of quantum evolution. Furthermore, the squeezing strength is chosen as $r=0.5$ and the QSL time shows symmetry about the squeezing phase $\theta=2.76$. From Fig.~\ref{QSL}(b), it is observed that the QSL time will increase with increasing the value of the squeezing strength before $r=0.18$ and then decrease. Without loss of generality, the squeezing phase is chosen as $\theta=\pi / 2$. Remarkably, we find that squeezed vacuum bath can affect the evolution of quantum system, thereby altering the speed of quantum correlation amplification.

\section{\label{Sec:6} CONCLUSIONS}
In conclusion, we have investigated the dynamics of quantum discord for two identical qubits in a common squeezed vacuum bath. Here, the $X \text {-type}$ state is selected as the initial state. According to the theoretical method, we have derived the equation for the critical time $t_c$, which is determined by the initial-state and reservoir parameters. Then we have numerically calculated the initial parameter intervals corresponding to finite and infinite critical time $t_c$, respectively. Moreover, we have studied the phenomenon of quantum discord amplification in these two different intervals.

In the case of finite critical time $t_c$, the squeezed vacuum reservoir with different squeezing parameters does not affect the maximum quantum discord of two identical qubits during the time evolution, which is mainly determined by the initial parameters of the system. Furthermore, the quantum discord is amplified over time before the abrupt transition occurs. However, the critical time can be efficiently controlled by adjusting squeezing parameters. More specifically, increasing the squeezing strength can prolong the time interval for decoherence-free subspace, while increasing the squeezing phase has the opposite effect. Even though the squeezing effect will disappear in the long-time evolution, the squeezed vacuum bath still plays an obvious regulatory role on the quantum discord amplification in the short-time limit. 

In the case of infinite critical time $t_c$, the quantum discord of two identical qubits coupled to a common squeezed vacuum reservoir can be amplified and its amplification will tend to be fixed in the long-time evolution. As in the case of finite critical time $t_c$, there also exists a decoherence-free subspace. Strikingly, different squeezing parameters result in different times to reach steady-state quantum discord. Besides, the degree of quantum discord amplification can be changed by adjusting the squeezing phase or squeezing strength in the decoherence evolution regime.

It has been indicated that a common squeezed vacuum reservoir may play a constructive or destructive role in the acceleration of the attainment of stable quantum discord, depending upon the squeezing parameters and the initial state of the two qubits. Here, the relative purity-based quantum speed limit bound is chosen to study the quantum speedup in the dynamic evolution of the system. Interestingly, we have revealed that the quantum speed limit time is influenced by reservoir squeezing and can be used as an effective method to measure the speed of reaching stable quantum correlation.

Finally, it should be noted that our theoretical results can be experimentally tested using cold atoms (impurities) subjected to a Bose-Einstein condensate environment. As shown in Refs.~\cite{Haikka2011,Addis2014}, the effective dephasing model with Ohmic like spectrum can be simulated. Furthermore, the squeezing effect on the bath can also be generated in Bose-Einstein condensates~\cite{Nir2011,Esteve2008}. Furthermore, squeezed reservoirs can be realized in a superconducting system~\cite{Murch2013}. Based on the experimental feasibility, squeezed vacuum reservoir technology is expected to be a new tool to control the generation of quantum discord and prolong the decoherence-free evolution time.

\acknowledgments This work was supported by the National Natural Science Foundation of China under Grants No.~12174346.


\begin{thebibliography}{99}


\bibitem{Yu2009} T. Yu and J. H. Eberly, Science \textbf{323}, 598 (2009).

\bibitem{Yu2004} T. Yu and J. H. Eberly, Phys. Rev. Lett. \textbf{93}, 140404 (2004).

\bibitem{Bellomo2007} B. Bellomo, R. Lo Franco, and G. Compagno, Phys. Rev. Lett. \textbf{99}, 160502 (2007).

\bibitem{de Moraes Neto2022} G. D. de Moraes Neto and V. Montenegro, Phys. Lett. A \textbf{438}, 128101 (2022).

\bibitem{Mintert2005} F. Mintert, A. R. R. Carvalho, M. Ku\'{s}, and A. Buchleitner, Phys. Rep. \textbf{415}, 207 (2005).

\bibitem{Chitambar2019} E. Chitambar and G. Gour, Rev. Mod. Phys. \textbf{91}, 025001 (2019).

\bibitem{Wootters1998} W. K. Wootters, Phys. Rev. Lett. \textbf{80}, 2245 (1998).

\bibitem{Vidal2002} G. Vidal and R. F. Werner, Phys. Rev. A \textbf{65}, 032314 (2002).

\bibitem{Horodecki2009} R. Horodecki, P. Horodecki, M. Horodecki, and K. Horodecki, Rev. Mod. Phys. \textbf{81}, 865 (2009).

\bibitem{Modi2010} K. Modi, T. Paterek, W. Son, V. Vedral, and M. Williamson, Phys. Rev. Lett. \textbf{104}, 080501 (2010).

\bibitem{Oppenheim2002} J. Oppenheim, M. Horodecki, P. Horodecki, and R. Horodecki, Phys. Rev. Lett. \textbf{89}, 180402 (2002).

\bibitem{Zurek2003} W. H. Zurek, Phys. Rev. A \textbf{67}, 012320 (2003).

\bibitem{Vedral2003} V. Vedral, Phys. Rev. Lett. \textbf{90}, 050401 (2003).

\bibitem{Maziero2009} J. Maziero, L. C. Céleri, R. M. Serra, and V. Vedral. Classical and quantum correlations under decoherence. Phys. Rev. A \textbf{80}, 044102 (2009).

\bibitem{Groisman2005} B. Groisman, S. Popescu, and A. Winter, Phys. Rev. A \textbf{72}, 032317 (2005).

\bibitem{Henderson2001} L. Henderson and V. Vedral, J. Phys. A \textbf{34}, 6899 (2001).

\bibitem{Bera2018} A. Bera, T. Das, D. Sadhukhan, S. Singha Roy, A. Sen(De), and U. Sen, Rep. Prog. Phys. \textbf{81}, 024001 (2018).

\bibitem{Datta2008} A. Datta, A. Shaji, and C. M. Caves, Phys. Rev. Lett. \textbf{100}, 050502 (2008).

\bibitem{Kaszlikowski2008} D. Kaszlikowski, A. Sen(De), U. Sen, V. Vedral, and A. Winter, Phys. Rev. Lett. \textbf{101}, 070502 (2008).

\bibitem{Ferraro2010} A. Ferraro, L. Aolita, D. Cavalcanti, F. M. Cucchietti, and A. Ac\'{i}n, Phys. Rev. A \textbf{81}, 052318 (2010).

\bibitem{Ollivier2001} H. Ollivier and W. H. Zurek, Phys. Rev. Lett. \textbf{88}, 017901 (2001).

\bibitem{Lanyon2013} B. P. Lanyon, P. Jurcevic, C. Hempel, M. Gessner, V. Vedral, R. Blatt, and C. F. Roos, Phys. Rev. Lett. \textbf{111}, 100504 (2013).

\bibitem{Fanchini2010} F. F. Fanchini, T. Werlang, C. A. Brasil, L. G. E. Arruda, and A. O. Caldeira, Phys. Rev. A \textbf{81}, 052107 (2010).

\bibitem{Maziero2010} J. Maziero, T. Werlang, F. F. Fanchini, L. C. C\'{e}leri, and R. M. Serra, Phys. Rev. A \textbf{81}, 022116 (2010).

\bibitem{Luo2008} S. Luo, Phys. Rev. A \textbf{77}, 042303 (2008).

\bibitem{Chen2011} Q. Chen, C. Zhang, S. Yu, X. X. Yi, and C. H. Oh, Phys. Rev. A \textbf{84}, 042313 (2011).

\bibitem{Ali2010} M. Ali, A. R. P. Rau, and G. Alber, Phys. Rev. A \textbf{81}, 042105 (2010).

\bibitem{Li2021} B. Li, C.-L. Zhu, X.-B. Liang, B.-L. Ye, and S.-M. Fei, Phys. Rev. A \textbf{104}, 012428 (2021).

\bibitem{Rulli2011} C. C. Rulli and M. S. Sarandy, Phys. Rev. A \textbf{84}, 042109 (2011).

\bibitem{Addis2015} C. Addis, G. Karpat, and S. Maniscalco, Phys. Rev. A \textbf{92}, 062109 (2015).

\bibitem{Haikka2013} P. Haikka, T. H. Johnson, and S. Maniscalco, Phys. Rev. A \textbf{87}, 010103(R) (2013).

\bibitem{ABasit2021} A. Basit, H. Ali, X.-F. Yang, and G.-Q. Ge, Phys. Scr. \textbf{96}, 075105 (2021).

\bibitem{Maziero2012} J. Maziero, L. C. C\'{e}leri, R. M. Serra, and M. S. Sarandy, Phys. Lett. A \textbf{376}, 1540 (2012).

\bibitem{Streltsov2017} A. Streltsov, G. Adesso, and M. B. Plenio, Rev. Mod. Phys. \textbf{89}, 041003 (2017).

\bibitem{Werlang2009}T. Werlang, S. Souza, F. F. Fanchini, and C. J. Villas Boas, Phys. Rev. A \textbf{80}, 024103 (2009).

\bibitem{Xu2011} L. Xu, J. B. Yuan, Q. S. Tan, L. Zhou, and L. M. Kuang, Eur. Phys. J. D \textbf{64}, 565 (2011).

\bibitem{Zhang2014} J.-S. Zhang and A.-X. Chen, J. Phys. B \textbf{47}, 215502 (2014).

\bibitem{Karpat2011} G. Karpat and Z. Gedik, Phys. Lett. A \textbf{375}, 4166 (2011).

\bibitem{Mazzola2010} L. Mazzola, J. Piilo, and S. Maniscalco,  Phys. Rev. Lett.  \textbf{104}, 200401 (2010).

\bibitem{mazzola2010} L. Mazzola, J. Piilo, and S. Maniscalco, Int. J. Quantum Inf. \textbf{09}, 981 (2010).

\bibitem{Poyatos1996} J. F. Poyatos, J. I. Cirac, and P. Zoller, Phys. Rev. Lett. \textbf{77}, 4728 (1996).

\bibitem{Murch2013} K. W. Murch, S. J. Weber, K. M. Beck, E. Ginossar, and I. Siddiqi, Nature \textbf{499}, 62 (2013).

\bibitem{Nir2011} N. Bar-Gill, D. D. B. Rao and G. Kurizki, Phys. Rev. Lett. \textbf{107}, 010404 (2011).

\bibitem{Akhtar2023} N. Akhtar, J. Wu, J.-X. Peng, W.-M. Liu, and G. Xianlong, Phys. Rev. A \textbf{107}, 052614 (2023).

\bibitem{Akhtar2022} N. Akhtar, B. C. Sanders, and G. Xianlong, Phys. Rev. A \textbf{106}, 043704 (2022).

\bibitem{Dey2020} S. Dey and S. S. Nair, J. Phys. A \textbf{53}, 385305 (2020).

\bibitem{Biswas2007} A. Biswas and G. S. Agarwal, Phys. Rev. A \textbf{75}, 032104 (2007).

\bibitem{Ablimit2023}  A. Ablimit, F.-H. Ren, R.-H. He, Y.-Y. Xie, and Z.-M. Wang, arXiv:2304.04223.

\bibitem{He2019}  Z. He, H.-S. Zeng, Y. Chen, and C. Yao, Laser Phys. Lett. \textbf{16}, 065204 (2019).

\bibitem{MAli2010}M. M. Ali, P.-W. Chen, and H.-S. Goan, Phys. Rev. A \textbf{82}, 022103 (2010).

\bibitem{You2018} Y.-N. You and S.-W. Li, Phys. Rev. A \textbf{97}, 012114 (2018).

\bibitem{Ro2014} J. Ro{\ss}nagel, O. Abah, F. Schmidt-Kaler, K. Singer, and E. Lutz, Phys. Rev. Lett. \textbf{112}, 030602 (2014).

\bibitem{Agarwalla2017} B. K. Agarwalla, J.-H. Jiang, and D. Segal, Phys. Rev. B \textbf{96}, 104304 (2017).

\bibitem{Assis2020} R. J. de Assis, J. S. Sales, J. A. R. da Cunha, and N. G. de Almeida, Phys. Rev. E \textbf{102}, 052131 (2020).

\bibitem{Assis2021} R. J. de Assis, J. S. Sales, U. C. Mendes and N. G. de Almeida, J. Phys. B \textbf{54}, 095501 (2021).

\bibitem{Klaers2017} J. Klaers, S. Faelt, A. Imamoglu, and E. Togan, Phys. Rev. X \textbf{7}, 031044 (2017).

\bibitem{Hernandez2008} M. Hernandez and M. Orszag, Phys. Rev. A \textbf{78}, 042114 (2008).

\bibitem{Naikoo2019} J. Naikoo, S. Banerjee, and A. M. Jayannavar, Phys. Rev. A \textbf{100}, 062132 (2019).

\bibitem{Drummond2020} R. Y. Teh, P. D. Drummond and M. D. Reid, Phys. Rev. Research \textbf{2}, 043387 (2020).

\bibitem{Banerjee2007} S. Banerjee and R. Ghosh, J. Phys. A \textbf{40}, 13735 (2007).

\bibitem{Wang2019} Y. Wang, C. Li, E. M. Sampuli, J. Song, Y. Jiang, and Y. Xia, Phys. Rev. A \textbf{99}, 023833 (2019).

\bibitem{Drummond2004}   P. D. Drummond and Z. Ficek, \emph{Quantum Squeezing} (Springer, Berlin, 2004).

\bibitem{Mandelstam1945} L. Mandelstam and I. Tamm, J. Phys. \textbf{9}, 249 (1945).

\bibitem{Margolus1998} N. Margolus and L. B. Levitin, Physica D \textbf{120}, 188 (1998).

\bibitem{Ashhab2012} S. Ashhab, P. C. de Groot, and F. Nori, Phys. Rev. A \textbf{85}, 052327 (2012).

\bibitem{Nie2021} S.-S. Nie, F.-H. Ren, R.-H. He, J. Wu, and Z.-M. Wang, Phys. Rev. A \textbf{104}, 052424 (2021).

\bibitem{Mukhopadhyay2018} C. Mukhopadhyay, A. Misra, S. Bhattacharya, and A. K. Pati,  Phys. Rev. E \textbf{97}, 062116 (2018).

\bibitem{D. P. Pires2021} D. P. Pires, K. Modi, and L. C. C\'{e}leri,  Phys. Rev. E \textbf{103}, 032105 (2021).


\bibitem{Giovannetti2011} V. Giovannetti, S. Lloyd, and L. Maccone, Nat. Photon. \textbf{5}, 222 (2011).

\bibitem{He2019} Z. He, H.-S. Zeng, Y. Chen, and C. Yao, Laser Phys. Lett. \textbf{16}, 065204 (2019).

\bibitem{Du2021} K. Y. Du, Y. J. Ma, S. X. Wu, C. S. Yu, Chin. Phys. B \textbf{30}, 090308 (2021).

\bibitem{Paulson2205} K. G. Paulson and S. Banerjee, arXiv:2205.11882.

\bibitem{Tiwari2023} D. Tiwari, K. G. Paulson, and S. Banerjee, Ann.
Phys. \textbf{535}, 2200452 (2023).

\bibitem{Paulson2022} K. G. Paulson and S. Banerjee, J. Phys. A \textbf{55}, 505302 (2022).

\bibitem{basit2021} A. Basit, H. Ali, F. Badshah, X.-F. Yang, and G.-Q. Ge, Laser Phys. Lett. \textbf{18}, 065202 (2021).

\bibitem{Basit2023} A. Basit, H. Ali, P.-B. Li, and G. Xianlong, Phys. Rev. A \textbf{107}, 042432 (2023).

\bibitem{Basit2021} A. Basit, H. Ali, F. Badshah, X.-F. Yang, and G. Ge, Phys. Rev. A \textbf{104}, 042417 (2021).

\bibitem{Yuan2010} J.-B. Yuan, L.-M. Kuang, and J.-Q. Liao, J. Phys. B \textbf{43}, 165503 (2010).

\bibitem{Yuan2013} J.-B. Yuan and L.-M. Kuang, Phys. Rev. A \textbf{87}, 024101 (2013).

\bibitem{Breuer2002} H. Breuer and F. Petruccione, \emph{The Theory of Open Quantum Systems} (Oxford University Press, Oxford, 2002).

\bibitem{Agarwal2012}   G. S. Agarwal, \emph{Quantum Optics} 1st ed.(Cambridge University Press, Cambridge, 2012).

\bibitem{Gardiner2004}   C. W. Gardiner and P. Zoller, \emph{Quantum noise: a handbook of Markovian and non-Markovian quantum stochastic methods with applications to quantum optics}  (Springer Science, 2004).

\bibitem{Chakravarty1984} S. Chakravarty and A. J. Leggett, Phys. Rev. Lett. \textbf{52}, 5 (1984).

\bibitem{Palma1996} G. M. Palma, K.-A. Suominen, and A. K. Ekert, Proc.
R. Soc. London A \textbf{452}, 567 (1996).

\bibitem{L. C.Celeri2009} J. Maziero, L. C. C\'{e}leri, R. M. Serra, V. Vedral, Phys. Rev. A \textbf{80}, 044102 (2009).

\bibitem{Deffner2017} S. Deffner and S. Campbell, J. Phys. A \textbf{50}, 453001 (2017).

\bibitem{Deffner2013} S. Deffner and E. Lutz, Phys. Rev. Lett. \textbf{111}, 010402 (2013).

\bibitem{Pires2016} D. P. Pires, M. Cianciaruso, L. C. Céleri, G. Adesso, and D. O. Soares-Pinto, Phys. Rev. X \textbf{6}, 021031 (2016).

\bibitem{Campaioli2018} F. Campaioli, F. A. Pollock, F. C. Binder, and K. Modi, Phys. Rev. Lett. \textbf{120}, 060409 (2018).

\bibitem{Wu2018} S-X. Wu and C-S. Yu, Phys. Rev. A \textbf{98}, 042132 (2018).

\bibitem{ZhangY2014} Y. J. Zhang, W. Han, Y. J. Xia, J. P. Cao, H. Fan, Sci. Rep. \textbf{4}, 4890 (2014).


\bibitem{Campo2013} A. del Campo, I. L. Egusquiza, M. B. Plenio, and S. F. Huelga, Phys. Rev. Lett. \textbf{110}, 050403 (2013).

\bibitem{S. Wu2014} S. Wu, Y. Zhang, C.-S Yu, and H. Song, J. Phys. A \textbf{48}, 045301 (2014).


\bibitem{Haikka2011} P. Haikka, S. McEndoo, G. De Chiara, G. M. Palma, and
S. Maniscalco, Phys. Rev. A \textbf{84}, 031602(R) (2011).

\bibitem{Addis2014} C. Addis, G. Brebner, P. Haikka, and S. Maniscalco, Phys. Rev. A \textbf{89}, 024101 (2014).

\bibitem{Esteve2008} J. Est\`{e}ve, C. Gross, A. Weller, S. Giovanazzi, and M. K. Oberthaler, Nature \textbf{455}, 1216 (2008).


\end{thebibliography}
\end{document}